\documentclass[pra ,showpacs,showkeys,twocolumn]{revtex4-1}

\usepackage{graphicx}
\usepackage{latexsym}
\usepackage{amsmath}
\usepackage{amssymb}
\usepackage{amsfonts}
\usepackage{color}
\usepackage{textcomp}
\usepackage{xfrac} 
\usepackage{physics}
\usepackage[normalem]{ulem}

\renewcommand{\a}[1]{\alpha_{#1}}
\renewcommand{\u}[1]{u_{#1}}
\renewcommand{\v}[1]{v_{#1}}
\newcommand{\dZ}[2]{\dd{Z}^{(#1)}_{#2}}

\DeclareMathOperator{\Oop}{\hat{O}}

\newcommand{\aop}{\hat{a}}
\newcommand{\cop}{{\aop^\dagger}}
\DeclareMathOperator{\delop}{\hat{\delta}}

\begin{document}

%\preprint{}

\title{Quantum trajectories for the variational description of closed systems:  
\\a case study with Gaussian states}

\author{Michiel Wouters}

\affiliation{TQC, Universiteit Antwerpen, Universiteitsplein 1,
B-2610 Antwerpen, Belgium}

\date{\today}

\begin{abstract}
It is proposed to improve the quality of a variational description of a closed quantum system by adding ficticious dissipation that reduces the entanglement. The proposal is implemented for a small Bose-Hubbard chain, which shows chaotic behavior and associated fast growth of quantum fluctuations. For appropriately chosen dissipation, good agreement with the truncated Wigner approximation (which is very accurate for the chosen system parameters) is found.
\end{abstract}

\maketitle

\section{Introduction}
Much progress in our understanding of many body quantum systems has come from the construction of variational wave functions, such the Hartree-Fock-Bogoliubov, Bardeen-Cooper-Schrieffer \cite{blaizot_ripka}, Laughlin \cite{quantumhall} and matrix product states \cite{mps}. For the description of ground states of a wide variety of physical systems, excellent variational wave functions are now available. The modelling of \textit{dynamics}, however, is still challenging for various many body systems. For example, MPS approaches often break down, because quick entanglement growth requires exponentially large computational resources \cite{prosen07}. Through the eigenstate thermalization hypothesis \cite{dalessio},  it is now well understood that closed nonintegrable quantum systems are at late times well described by a Boltzmann-Gibbs state
\begin{equation}
|\psi(t) \rangle \sim \rho_{BG} \propto e^{-\beta H}, \;\; \textrm{for}\; t \rightarrow\infty
\end{equation} 
where the `$\sim$'-sign means that the states are (approximately) equivalent for what concerns the local observables. Even though the true eigenstates are highly entangled, the density matrix can be well approximated by a mixture of states with low entanglement \cite{stoudenmire10}.

At intermediate times, the situation is the most difficult, because entanglement has already grown significantly, while the Gibbs state has not yet been reached. 
In order to describe the system in the intermediate regime, inspiration can be taken from the equilibrium state. If one would describe the latter as a pure state, it would also be highly entangled. It then seems a reasonable assumption that also the large entanglement states at intermediate times can be approximately represented by a classical mixture of low-entanglement states $|\phi_j(t) \rangle$:
\begin{equation}
|\psi(t) \rangle \sim \sum_j p_j |\phi_j(t) \rangle \langle \phi_j(t) |.
\label{eq:apprmix}
\end{equation}

While the time evolution of a state in terms of a single wave variational wave function can be computed with the time-dependent variational principle (TDVP), to the best of my knowledge no such powerful principle exists for the approximate representation \eqref{eq:apprmix}. In the context of open quantum systems, the evolution from a pure to a mixed state of the form \eqref{eq:apprmix} is obtained in the quantum trajectory approach \cite{carmichael93,breuer}. The increasing mixedness of the state is then due to the uncertainty in the outcome of the measurements that are performed on the system. Due to the interaction of the system with the environment, the entanglement can be greatly reduced \cite{nha04,chan19,skinner19,li18,li19} and the trajectory wave functions are more easily described by a variational wave function.

%\textit{Proposal -- } 

I will argue here that for the dynamics of closed systems, the range of applicability of variational wave functions can be improved by adding jump operators to the dynamics. In other words, the variational description of the closed system with added jump operators can be closer to the real system dynamics than the variational description of the system itself.
The intuition is that the coherence between macroscopically different states 
(Schr\"odinger cats) cannot be physically important (i.e. very hard to measure and therefore in macroscopic systems actually `unphysical'). Opening the system by including suitably chosen jump operators then explicitly eliminates these unphysical coherences without disturbing the system much. Not disturbing the system at all is impossible, but at the same time, it is also impossible to exactly represent the state by the variational ansatz. My claim is that the disturbance due to the measurement can be less than its resulting improvement of the variational description. 

A complementary way to see the that the effect of additional jump operators can be negligible is from the classical chaos in the TDVP equations of motion.
The growth of entanglement is directly related to chaos, both in a Gaussian \cite{bianchi18} and MPS \cite{hallam19} variational description.
In the chaotic case, the exponential growth of the fluctuations and bond dimension respectively make it better to describe the state as a mixture of variational states, which can be achieved through the quantum trajectory approach. The price to be paid is a deviation from the the exact dynamics, but in the case of chaotic dynamics, the resulting deviation is swamped by the exponential growth of fluctuations due to the nonlinear dynamics.

This proposal is related to the purification of density matrices with matrix product states in a Hilbert space enlarged with an ancilla \cite{verstraete04,barthel09}, where the ancilla plays the role of environment. In the present approach, no ancilla is included, but the environment is introduced through Lindblad jump operators.

\section{Example: Bose-Hubbard chain}
As a toy system to gain more insight in the above claim, I will consider a small Bose-Hubbard chain, described by the Hamiltonian
\begin{equation}
H= \sum_{j=1}^N \left( -J
\hat a_{j+1}^\dag \hat a_j   -J a_j^\dag \hat a_{j+1} 
+\frac{U}{2} \hat a_j^\dag \hat a_j^\dag \hat a_j \hat a_j 
\right).
\end{equation}
The variational description will be made in terms of Gaussian wave functions, which becomes accurate in the mean field limit. It corresponds to a Wick decoupling of the equations of motion for the first and second order correlation functions (see appendix \ref{sec:eqGVA}), yielding the so-called Hartree-Fock-Bogoliubov (HFB) description.

The coupling to the bath will be taken in the form of heterodyne detection of single particle losses \cite{breuer}. Physically, this is done by interfering the bosons that are lost from the coupling to the environment with a detuned local oscillator. This measurement scheme gives information on both the amplitudes and phases of the Bose fields. The disadvantage for our purposes is that it involves the loss of particles, which strongly impacts the dynamics. In order to avoid this discrepancy between the open system and the closed system of interest, the state is projected back onto the manifold with the initial particle number. Any type of conservation law that is violated by the opening of the system can be restored in this way. Since we use the opening of the system as a mathematical tool to improve the variational description, it is not important that a physical implementation exists, but it could be achieved by introducing feedback \cite{wiseman}.

The advantage of the Bose Hubbard system is that it is in the mean field limit up to a very good approximation described by the truncated Wigner approximation (TWA) \cite{sinatra02}, as illustrated in appendix \ref{sec:TWA}. Within the TWA, the quantum fields are replaced by classical fields, that obey the Gross-Pitaevskii equation (GPE)
\begin{equation}
i\frac{\partial}{\partial t} \alpha_j=-J\left(\alpha_{j-1}+\alpha_{j+1}\right)+U |\alpha_j|^2 \alpha_j
\end{equation}
The quantum origin is reflected in the stochastic initial conditions. For an initial coherent state with amplitudes $\alpha^{(0)}_j$, the $\alpha_j$ have a Gaussian distribution, centered at $\alpha^{(0)}_j$ and   variance $\sfrac{1}{2}$. 
With deterministic initial condition, the GPE describes the variational dynamics within the manifold of coherent states, which is a submanifold of the Gaussian states that is used here.

\begin{figure} \centering
\includegraphics[width=0.9\linewidth]{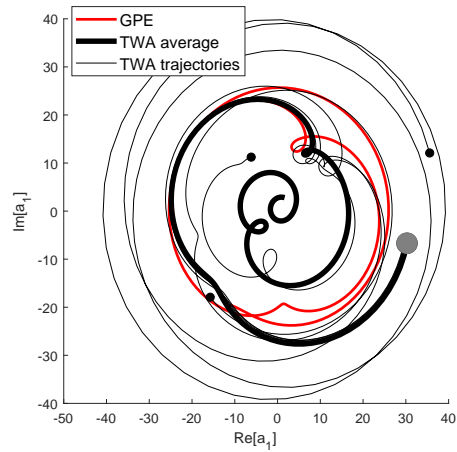}
\caption{ Illustration of the chaos in the Bose-Hubbard system. The TWA average strongly differs from the GPE because the trajectories with slightly different initial conditions evolve very differently. Black dots visualize the end points of the TWA trajectories, the gray dot indicates the initial condition. Parameters: total particle number $N=3090$, and $U/J=0.01$ and initial condition 
$\vec{\alpha}_0=(30-6.6i, 9.7-7.3i, 40,20)$.
} \label{fig:1}
\end{figure}

The Bosonic Hubbard model has two conserved quantities, the energy and number of particles, such that it is integrable for one and two sites. From three sites on, it starts to show chaos \cite{cruzeiro90,kidd17}. For the TWA description, this implies that trajectories for the initially slightly different initial conditions quickly diverge \cite{kidd17}. An illustrative example with four sites is shown in Fig. \ref{fig:1}. The amplitude at the first site for a coherent state initial condition is shown in the complex plane as obtained with the TWA (thick black line). The red line shows the GPE evolution of the initial state; the thin black lines show a few other TWA trajectories. The black dots indicate the very different final amplitudes of the TWA trajectories.

\begin{figure} \centering
\includegraphics[width=0.85\linewidth]{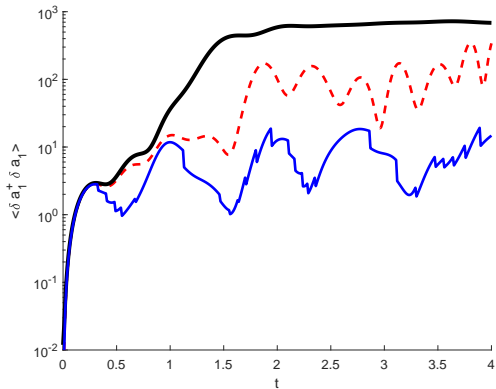}
\caption{ Growth of fluctuations on the first site computed with TWA (thick black line), with Gaussian varational wave function (red dashed line) and in the TGA with $\delta N_{\rm max}=20$ (blue line). Same parameters as in Fig. \ref{fig:1} (a).} 
\label{fig:2}
\end{figure}

When starting from a coherent state, at early times the fluctuations are small and both the TWA and Gaussian variational approaches are very accurate. Only when the fluctuations become large, a difference between the TWA and the variational Gaussian descriptions appears. As an illustration, Fig. \ref{fig:2} shows the growth of fluctuations on the first site as a function of time. The chaotic nature of the system is reflected in the quick growth of fluctuations, up to their maximal value determined by the finite available phase space volume, that is limited by energy and number conservation. The TWA (tick black line) and Gaussian variational (dashed red line) calculations coincide at early times, when fluctuations are small. Deviations appear when the nonlinear effects, incompletely captured by the Gaussian approximation, become important.

The usefulness of including a fictitious homodyne detection in the dynamics comes precisely from the fact that it reduces the fluctuations as can be seen from the TGA equations given in appendix \ref{sec:eqGVA}. Omitting for clarity the terms due to the Hamiltonian dynamics, they reduce to
\begin{align}
\dd \a{n}&=-\frac{\gamma}{2}\a{n} \dd{t}+\sqrt{\gamma}\sum_i\left(\v{in}\dZ{1}{i}+\u{in}{\dZ{1}{i}}^*\right), \label{eq:dameas} \\
\dd \u{nm}&= -\gamma \left[\u{nm} + \sum_i\left(\u{mi}\v{in}+\u{ni}\v{im}\right)\right] \dd{t}, \label{eq:dumeas} \\
\dd \v{nm}&=-\gamma \left[\v{nm} +\sum_i\left(\v{ni}\v{im}+\u{ni}^*\u{im}\right)\right]\dd{t}.
\label{eq:dvmeas}
\end{align}
where
$\alpha_n=\ev*{\aop_n}$, $u_{nm}:=\ev*{\delop_n\delop_m}=\ev*{\aop_n\aop_m}-\a{n}\a{m}$, $v_{nm}:=\ev*{\delop_n^\dagger\delop_m}=\ev*{\cop_n\aop_m}-\a{n}^*\a{m}$. The noise term
$dZ_i=\frac{1}{\sqrt{2}}(dW_{x,i}+idW_{p,i})$ is a complex Wiener process satisfying 
$\abs{dZ_i}^2=\dd{t}$. 
 
Eqs. \eqref{eq:dumeas} and \eqref{eq:dvmeas} show that the measerument at rate 
$\gamma$ reduces the fluctuations. Eq. \eqref{eq:dameas} has a deterministic contribution expressing a loss of particles and a stochastic term, that transfers the reduction of the fluctuations described by Eqs. \eqref{eq:dumeas} and \eqref{eq:dvmeas}. The deterministic loss term in Eq. \eqref{eq:dameas} reduces the number of particles -- a conserved quantity under the Hamiltonian dynamics -- leading to a discrepancy between the closed and open systems. As discussed above, the state after infinitesimal time evolution can be projected back onto the manifold with the correct particle number. As a consequence of this projection, the first term on the r.h.s. of Eq. \eqref{eq:dameas} is removed.

Ito calculus with the stochastic part of the dynamics of Eq. \eqref{eq:dameas} yields for the realization averaged correlation function:
\begin{equation}
d \ev{\alpha_n \alpha_m} = \gamma \sum_i \left\langle \u{mi} \v{in}  + \u{ni} \v{im} \right\rangle \dd{t}.
\label{eq:daa}
\end{equation}
The total second moment of of the annihilation operators is
\begin{equation}
\ev{\hat a_n \hat a_m} = \ev{\alpha_n \alpha_m }+  \ev{\u{mn}},
\label{eq:aa_tot}
\end{equation}
 In the dynamics of the total second moment \eqref{eq:aa_tot}, the contribution \eqref{eq:daa} is canceled by by the second term in the square brackets in Eq. \eqref{eq:dumeas}. Under the condition that the latter term dominates the first one in Eq. \eqref{eq:dumeas}, which is the case when the magnitude of the fluctuations is much larger than one, the total second moment is to a good approximation unaffected by dissipative dynamics. The dynamics of the other second moment $\ev{\hat a^\dag_m \hat a_n}$ is fully analogous.

\begin{figure} \centering
\includegraphics[width=0.9\linewidth]{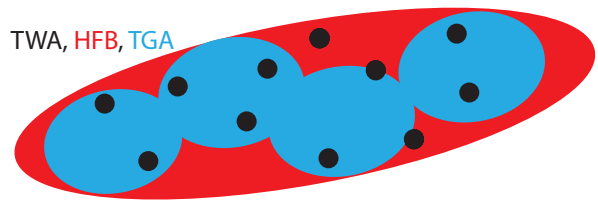}
\caption{Sketch of the different representations of the state in the TWA, HFB and TGA. In truncated Wigner, points in phase space are sampled, illustrated by the black dots. In the HFB, a single squeezed state is evolved, as represented by the red ellipse. In the TGA, the state is represented by a mixture of squeezed states.} \label{fig:compare}
\end{figure}

In the quantum trajectory description, quantum (intratrajectory) fluctuations are then converted into classical (intertrajectory) fluctuations, keeping the total amount of fluctuations constant. A pictorial illustration of this mechanism  is presented in Fig. \ref{fig:compare}. The black dots represent the TWA sampling of the state, the red ellipse the squeezed Gaussian state in HFB and the blue ellipses the ensemble of squeezed states in the TGA. The TGA is a resampling of the HFB, which reduces the fluctuations and therefore improves the quality of the Gaussian representation.

From the above considerations, it appears that dissipation should be strong enough to keep the fluctuations sufficiently small for the Gaussian approximation to be valid, but at the same time the dissipation should be weak enough so that the fluctuations are sufficiently large for the total fluctuations to be only weakly affected. 

In order to study the dependence of the TGA on the strength of dissipation, a two-step dynamics was implemented, with separated Hamiltonian and dissipative evolutions. The Hamiltonian evolution is performed until the on-site flucutations exceed on any site a certain threshold $\delta N_{\rm max}$. Then, a dissipative evolution starts until the fluctuations are suppressed to $\delta N_{\rm min}$, which was taken $\delta N_{\rm min}=\delta N_{\rm max}/2$. The evolution of the fluctuations in this approach is shown by the blue line in Fig. \ref{fig:2}. The sudden downward jumps in the fluctuations are due to the dissipative evolution. 

\begin{figure} \centering
\includegraphics[width=1.\linewidth]{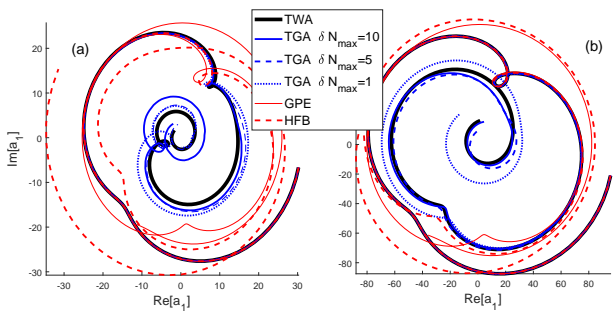}
\caption{Comparison of the TWA, GPE and TGA with various maximal fluctuation numbers for a four-site Bose-Hubbard model. Parameters: (a) total particle number $N=3090$, and $U/J=0.01$; (b) $N=30900$, $U/J=0.  001$.
 } \label{fig:34}
\end{figure}

Numerical examples for the expecation value of the amplitude on the first site are shown in Fig. \ref{fig:34}. From the numerics, it is clear that the TGA lies much closer to the TWA than the GPE or HFB. The dependence on the fluctuation threshold for dissipation $\delta N_{\rm max}$ turns out to be weaker than expected: also the simulation with small $\delta N_{\rm max}$ are close to the TWA. 
One does see that deeper in the mean field limit (panel (b)), the lower threshold performs notably worse than the higher ones, which does agree with the analytical arguments.

%\begin{figure} \centering
%\includegraphics[width=0.9\linewidth]{fig3.eps}
%\caption{Comparison of the TWA, GPE and TGA with various maximal fluctuation numbers for a four-site Bose-Hubbard model. Parameters: total particle number $N=400$, $U/J=0.01$. } \label{fig:3}
%\end{figure}
%
%\begin{figure} \centering
%\includegraphics[width=0.9\linewidth]{fig4.eps}
%\caption{Comparison of the TWA, GPE and TGA with various maximal fluctuation numbers for a four-site Bose-Hubbard model. Parameters: total particle number $N=4000$ and $U/J=0.001$.} \label{fig:4}
%\end{figure}

\section{Discussion}
The most important lesson that follows from the Bose-Hubbard example is the fact that the variational trajectory description with fictitious dissipation describes the chaotic closed system much better than the straightforward variational description. The role of classical chaos is clear: in the absence of chaos, the Gaussian fluctations remain small in the semiclassical limit, the usual variational principle is then accurate and adding dissipation will make the description only worse.

For the present case study, this insight does not come with a technical advantage, because the truncated Wigner approach works even better (see Appendix \ref{sec:TWA}) and is also computationally less expensive. The efficiency of the TWA however depends on the fact that the system under consideration is close to the classical field limit. For systems that do not have such a close classical counterpart and need a more complicated variational description such as MPS, the present proposal of adding fictitious dissipation could give theoretical descriptions of systems that are currently out of reach.

What is shared by both the Gaussian and MPS states is that they perform poorly for highly entangled states, the Gaussian states because they are a bad approximation, the MPS because they consume prohibitively large computational resources. The advantage of adding dissipation is that it reduces the entanglement. For the Gaussian state, this is readily seen from Eqs. \eqref{eq:dumeas}, \eqref{eq:dvmeas} together with the fact that entanglement scales with the fluctuation magnitude \cite{serafini}. 
With appropriately chosen jump operators and unravelling scheme, this is expected to be also the case for other variational states. 

From the Gaussian example, it is also seen that the flexibility in the unraveling can be exploited to improve the accuracy of the variational trajectory approach. For single-particle losses, there are two common unraveling schemes: particle counting and homo/heterodyne detection  \cite{nha04}. Under particle counting, no phase information is retrieved and Schr\"odinger cats of coherent states are still present. For this detection scheme, the trajectories are therefore far from Gaussian. For heterodyne detection on the other hand, the individual trajectory states have a well defined phase and the Gaussian approximation is satisfactory \cite{verstraelen18}.

Conceptually, also the TWA method can actually be described in the terms similar to the trajectory approach. In the TWA, coherent states that are close -- but not identical -- to the initial state are variationally evolved. In the presence of chaos, this time evolution leads to very different states. A mixture of these is a much better approximation to the actual state of the system than the variational evolution of the initial state only. 
This could give a complementary view on the nature of TWA, which is rigorously motivated as an approximation to the evolution of the phase space quasi-probability evolution \cite{quantumnoise}.
The TWA approach for fermions from Ref. \cite{davidson17} could also be interpreted from this perspective.
It would be interesting to see whether the evolution of an ensemble of variational states sampled around the initial state could be generalized to other types of variational wave functions.

The importance of the decomposition \eqref{eq:apprmix} is that it preserves the notion of locality, which is lost in the exact time evolution. Therefore, the variational description is in a sense more physical than the exact description, in the same way that this is the case for systems with spontaneous symmetry breaking \cite{wen}. For example in a transverse field field ferromagnet, the ground state is a GHZ like superposition of the states with both orientations of the spins. The lack of robustness of this state with respect to local decoherence however makes it unphysical. The physical states can be obtained from the exact ground state by measuring a single spin.

With the insights from this analysis, the dynamics can be divided in three stages: i) an initial low entanglement stage, ii) a large entanglement stage for which no description with a single local state exists, iii) thermal equilibrium, that is fully characterized by the (local) Hamiltonian and the energy. It is in the second phase that the system shows the most complex behavior. It is the phase of the dynamics where the state cannot be well approximated by a density matrix of the form
\begin{equation}
\rho = e^{-H_E},   \;\textrm{with} \; H_E \; \textrm{a \textit{local} operator.}
\end{equation}
Without the restriction that the entanglement hamiltonian $H_E$ is a local operator, the system can always be well approximated by such a state, but then it does not give any physical insight. Instead, one can give a description of the form
\begin{equation}
\rho \approx \sum_{k=1}^\mathcal{C}  p_k\; e^{- H_E^{(k)}},
\label{eq:rho_dec}
\end{equation}
where the entanglement Hamiltonians $H_E^{(k)}$ are required to be local.
With the Gaussian trajectory approach, this is achieved: the density matrix is written as a sum over approximately coherent states, which correspond to the entanglement Hamiltonian $H^{(k)}_E \propto \sum_i(a_i^\dag -\alpha^{(k)*}_i)(a^{(k)}_i -\alpha_i)$, where the indices $i$ and $k$ refers to the lattice sites and trajectories respectively. The trajectory approach makes it possible to study the growth of the number of components $\mathcal{C}$ in the early stages of the dynamics. At late times, $\mathcal{C}$ should decrease, because the system approaches the Boltzmann-Gibbs state. The reduction at later times can however not be so simply captured by the trajectory approach and is therefore beyond the scope of this discussion.

\section{Conclusions}
I have shown that adding dissipation to a closed quantum system can improve the variational description of the dynamics in the chaotic regime. As a case study, a small Bose-Hubbard chain was described by a Gaussian variational state. Adding heterodyne detection of the bosons improves the quality of this variational description notably when particle losses are eliminated by projecting back on the number conserving variational manifold. Adding dissipation for the improvement of variational descriptions could also be useful for other types of variational wave functions. 

\section{Acknowledgements}
Stimulating discussions with Wouter Verstraelen and Dolf Huybrechts are warmly acknowledged. This work was supported by the Research Foundation - Flanders (FWO) through the grant FWO-39532.

\appendix

\begin{widetext}

\section{equations of motion for TGA \label{sec:eqGVA}}

Following the approach of \cite{verstraelen19}, the evolution of a generic expectation value $\ev*{\Oop}$ under heterodyne unraveling for all decay channels is given by
\begin{align}\label{eq:generictrajectorycorrelation}
    \dd\ev*{\Oop}=&i\ev{\comm{\hat{H}}{\Oop}}\dd{t}-\frac{1}{2}\sum_{j,k}\left(\ev{\acomm{\hat{\Gamma}_{j,k}^\dagger \hat{\Gamma}_{j,k}}{\Oop}}-2\ev{\hat{\Gamma}_{j,k}^\dagger\Oop\hat{\Gamma}_{j,k}}\right)\dd{t}\nonumber\\&+\sum_{j,k}\left(\ev{\hat{\Gamma}_{j,k}^\dagger(\Oop-\ev*{\Oop})}\dd{Z}_{j,k}+\ev{(\Oop-\ev*{\Oop})\hat{\Gamma}_{j,k}}\dd{Z}_{j,k}^*\right),
\end{align}
where $dZ_i=\frac{1}{\sqrt{2}}(dW_{x,i}+idW_{p,i})$ is a complex Wiener process satisfying $\abs{dZ_i}^2=dt$. By assuming a Gaussian ansatz, Wick decompositions are performed and and the trajectory is expressed entirely in terms of the first and second central moments $\alpha_n=\ev*{\aop_n}$, $u_{nm}:=\ev*{\delop_n\delop_m}=\ev*{\aop_n\aop_m}-\a{n}\a{m}$, $v_{nm}:=\ev*{\delop_n^\dagger\delop_m}=\ev*{\cop_n\aop_m}-\a{n}^*\a{m}$. Note that $\u{nm}=\u{mn}$ and $\v{nm}=\v{mn}^*$. Using \eqref{eq:generictrajectorycorrelation}, the evolution of Gaussian moments is given by

\begin{align}\label{eq:gausseq}
\dd \a{n}&=\left[\left(-\frac{\gamma}{2}+i\Delta\right)\a{n}+i\frac{J}{z}\sum_{n'}\a{n'}
-iU(\abs{\a{n}}^2\a{n}+2\a{n}\v{nn}+\a{n}^*\u{nn})\right]\dd{t}\nonumber\\
&+\sqrt{\gamma}\sum_i\left(\v{in}\dZ{1}{i}+\u{in}{\dZ{1}{i}}^*\right)\\
\dd \u{nm}&=\left[2i\Delta\u{nm}+i\frac{J}{z}\left(\sum_{n'}\u{n'm}+\sum_{m'}\u{nm'}\right)
\right.\nonumber\\
&\left.-iU \left(\v{nm}(\a{n}^2+\u{nn})+\v{mn}(\a{m}^2+\u{mm})+2\u{nm}(\abs{\a{n}}^2+\abs{\a{m}}^2+\v{nn}+\v{mm})+\delta_{n,m}(\a{n}\a{m}+\u{nm})\right)\nonumber\right.\\
&\left.-\gamma\sum_i\left(\u{nm}+\u{mi}\v{in}+\u{ni}\v{im}\right)\right] \dd{t} \\
\dd \v{nm}&=\left[iU\left(2\v{nm}(\abs{\a{n}}^2-\abs{\a{m}}^2+\v{nn}-\v{mm})+\u{nm}(\a{n}^{*2}+\u{nn}^*)-\u{nm}^*(\a{m}^2+\u{mm})\right)-i\frac{J}{z}\left(\sum_{n'}\v{n'm}-\sum_{m'}\v{nm'}\right)
\right.\nonumber\\
&\left.-\gamma \v{nm}-\gamma\sum_i\left(\v{ni}\v{im}+\u{ni}^*\u{im}\right)\right]\dd{t}\nonumber\\
\end{align}
where primed indices refer to nearest-neighbours only.

\end{widetext}

\section{Comparison between TWA and exact diagonalization \label{sec:TWA}}

\begin{figure} \centering
\includegraphics[width=\linewidth]{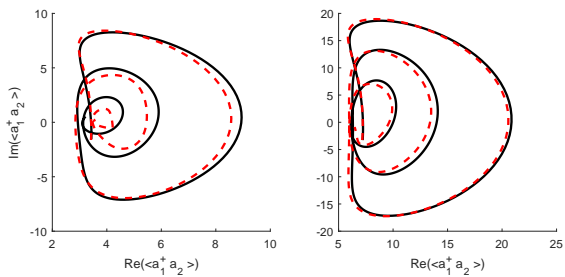}
\caption{Comparison of the TWA (red dashed line) with exact diagonalization (black full line) for a three site Bose-Hubbard model. Parameters: (a) total particle number $N=30$, $U/J=0.5$ and (b) $N=60$, $U/J=0.25$. } \label{fig:exact}
\end{figure}

The TWA is a good approximation in the semiclassical limit, where $N \rightarrow \infty, U \rightarrow 0, UN=c^{te}$. For small systems (we will use here three cavities), the time evolution can still be evaluated by exact diagonalization of the Hamiltonian, where particle number conservation can be used to reduce the size of the Hilbert space.

The comparison in Fig. \ref{fig:exact} of both approaches gives an indication of the reliability of the TWA. For this example, there are on average 10 particles per site (panel (a)) and 20 per site (panel (b)). In both cases, the agreement is satisfactory, and it is also clear that it improves deeper in the mean field limit (panel (b)).

\end{document}